\pdfoutput=1
\documentclass{article}
\usepackage{arxiv}
\usepackage[utf8]{inputenc} 
\usepackage[T1]{fontenc}    
\usepackage{hyperref}       
\usepackage{url}            
\usepackage{booktabs}       
\usepackage{amsfonts}       
\usepackage{nicefrac}       
\usepackage{microtype}      
\usepackage{subfig}
\usepackage{svg}
\usepackage{lipsum}
\usepackage{fancyhdr}       
\usepackage{graphicx}       
\usepackage{pifont}
\newcommand{\cmark}{\ding{51}}
\newcommand{\xmark}{\ding{55}}
\usepackage{multirow}
\usepackage{pdflscape}
\usepackage{rotating}
\usepackage{caption}
\graphicspath{{media/}}     
\hypersetup{
    colorlinks=true,
}

\title{\emph{\textquotesingle Aariz}: A Benchmark Dataset for Automatic Cephalometric Landmark Detection and CVM Stage Classification}

\author{
    {Muhammad Anwaar Khalid}\textsuperscript{1} \\
    \And
    {Kanwal Zulfiqar}\textsuperscript{2} \\
    \And
    {Ulfat Bashir}\textsuperscript{2} \\ 
    \And
    {Areeba Shaheen}\textsuperscript{2} \\
    \And
    {Rida Iqbal}\textsuperscript{2} \\
    \And
    {Zarnab Rizwan}\textsuperscript{2} \\
    \And
    {Ghina Rizwan}\textsuperscript{2} \\
    \And
    {Muhammad Moazam Fraz}\textsuperscript{1} \\
    \And
    \textnormal{
        \textsuperscript{1}{National University of Sciences and Technology (NUST), Islamabad, Pakistan}} \\
    \textnormal{
        \textsuperscript{2}{Riphah International University, Islamabad, Pakistan}} \\
}

\begin{document}
\maketitle
    
    \begin{abstract}
    The accurate identification and precise localization of cephalometric landmarks enable the classification and quantification of anatomical abnormalities. The traditional way of marking cephalometric landmarks on lateral cephalograms is a monotonous and time-consuming job. Endeavours to develop automated landmark detection systems have persistently been made, however, they are inadequate for orthodontic applications due to unavailability of a reliable dataset. We proposed a new state-of-the-art dataset to facilitate the development of robust AI solutions for quantitative morphometric analysis. The dataset includes 1000 lateral cephalometric radiographs (LCRs) obtained from 7 different radiographic imaging devices with varying resolutions, making it the most diverse and comprehensive cephalometric dataset to date. The clinical experts of our team meticulously annotated each radiograph with 29 cephalometric landmarks, including the most significant soft tissue landmarks ever marked in any publicly available dataset. Additionally, our experts also labelled the cervical vertebral maturation (CVM) stage of the patient in a radiograph, making this dataset the first standard resource for CVM classification. We believe that this dataset will be instrumental in the development of reliable automated landmark detection frameworks for use in orthodontics and beyond.
    \end{abstract}
    

    \section{Introduction}
    Quantitative morphometry of human skull and cephalometric analyses of spatial relationships among teeth, jaws, and cranium are considered to be the cornerstones of contemporary orthodontics, orthognathic treatment, and other areas of oral and maxillofacial surgeries \cite{proffit2006contemporary}. Such analyses are usually carried out using two-dimensional (2D) radiographic images often referred to as cephalograms. In 1982, Thomas Rakosi defined 90 anatomically relevant anchor points, i.e., landmarks, of which 29 have been widely used by orthodontists on a routine basis \cite{rakosi1982atlas}. A conventional cephalometric analysis involves the identification of anatomical landmarks, measurement of various angles and distances between these identified landmarks, and qualitative assessment of anatomical abnormalities from these angles and distances \cite{arik2017fully}. It assists clinicians in diagnosing the craniofacial condition of a patient by providing clinical interpretation of bony structures of the skull and surrounding soft tissues. In clinical practice, orthodontists usually map out contours of craniofacial structures on X-ray images manually and then identify anatomical landmarks from various geometrical features e.g. vertical and horizontal reference lines etc. However, the process of manual annotation is tedious, time-consuming and subjective \cite{kamoen2001clinical}. Although cephalometric tracing is generally performed by trained orthodontists, several reports have raised concerns regarding significant inter- and intra-observer variabilities among them \cite{durao2015cephalometric} due to their diverse training and experience backgrounds. Since identifying anatomical abnormalities of hard and soft tissues and subsequent treatment procedures are highly sensitive to precise estimation of landmark locations, a poor manual cephalometric analysis might have severe repercussions. Therefore, to improve the accuracy and reliability of cephalometric tracing, fully automated landmark detection systems have been in a long-standing area with great demand.
    
    In recent years, several research studies \cite{chen2019cephalometric, zhong2019attention, lee2020automated, zeng2021cascaded, wang2021dcnn, he2021cephalometric} have been proposed on computer-aided localization of anatomical landmarks and soft tissue boundaries. However, a comprehensive review of literature to date revealed that the automated cephalometric landmark detection systems have only achieved a success detection rate of 82.03\% within 2.0 mm, which is the clinically accepted precision range for landmark prediction. This is primarily due to the lack of adequate landmark tracing datasets for cephalometric analysis, as the annotation process is expensive and involves patients' privacy concerns. One of the main limitations of currently available datasets for automated cephalometric landmark detection is the insufficient number of training images, making it difficult for AI models to be trained effectively. Additionally, following the soft-tissue paradigm, several soft-tissue landmarks that are frequently used in clinical practice today, are not available in these datasets as well. Furthermore, the use of images from a single imaging device in these datasets has resulted in a tendency for trained AI models to over-fit and demonstrate sub-optimal performance on test data \cite{arik2017fully, lee2020automated, zeng2021cascaded} and in real settings. Therefore, a dataset with a large number of training images, acquired from different imaging devices with varying resolutions, and annotated with the most commonly used anatomical landmarks, is still needed.

    In addition to selecting the most appropriate treatment approach and biomechanics, the timing of treatment delivery is also crucial for orthodontic treatment effectiveness. Accurate assessment of facial growth and quantification of skeletal maturation stage, particularly mandibular growth, plays a vital role in orthodontic diagnosis, prognosis, treatment planning and outcomes. If the treatment is initiated at optimal developmental phase, it might result in more favourable outcomes. Otherwise, the treatment duration will be prolonged or surgical intervention might become necessary to rectify the jaw deformities \cite{hunter1966correlation} \cite{mcnamara2018cervical}. Several biological indicators, such as dental development and eruption times \cite{franchi2008phases}, chronological and dental age \cite{fishman1979chronological}, ossification of hand and wrist bones \cite{flores2004use}, and cervical vertebral maturation (CVM) and morphology \cite{baccetti2005cervical}, have been proposed to assess skeletal maturity. Among these, the CVM stage method is the most commonly used by orthodontists since it can be assessed using traditional lateral cephalograms and avoids additional radiographic exposure of a second radiograph. Despite the widespread use of the CVM stages method in assessing skeletal maturation and growth spurts, the process can be challenging and time-consuming for practitioners, leading to sub-optimal intra-observer agreement \cite{zhao2012validity}. This highlights the difficulty in accurately assessing CVM stages, which can be addressed through the use of qualitative assessment techniques or intelligent systems \cite{nestman2011cervical}. Over the years, several researchers \cite{kok2019usage, makaremi2019deep, amasya2020validation, seo2021comparison, atici2022fully, mohammad2022deep} have explored the promising applications of artificial intelligence (AI) to automate the process of CVM assessment, however, it is quite surprising that there remains a lack of a standard dataset for CVM stage estimation.
    
    Considering all these limitations, we propose a new benchmark dataset consisting of 1000 cephalometric X-ray images acquired from 7 different X-ray imaging devices with varied resolutions. The dataset includes annotations for two different cephalometric analyses: anatomical landmark detection and CVM stage classification. A team of 6 clinicians, including 2 expert orthodontists, were responsible for the annotation process, which was carried out in two phases. In the first phase, referred to as the labelling phase, two junior orthodontists independently annotated all of the cephalograms with 29 most commonly used anatomical landmarks. In the second phase, referred to as the reviewing phase, two senior orthodontists collaboratively reviewed and corrected the annotations. For CVM stage classification, our senior orthodontists assigned each image to one of the six possible stages, and the images for which they disagreed, were referred to expert orthodontists. The salient features of our dataset are summarized as follows:
    
    \begin{itemize}
        \item Our dataset boasts a diverse and extensive collection of 1000 cephalograms acquired from 7 different X-ray imaging devices with varying resolutions, making it the most comprehensive cephalometric dataset to date.
        \item The dataset features 29 most commonly used anatomical landmarks, with 15 skeletal, 8 dental, and 6 soft-tissue landmarks, annotated by a team of 6 skilled orthodontists in two phases, following extensive labelling and reviewing protocols.
        \item By annotating the CVM stages of each cephalogram in our dataset, we have also created the first standard resource for automatic CVM classification.
    \end{itemize}
    
    These aspects make our dataset a new state-of-the-art in the field of quantitative cephalometry. We named our dataset \emph{\textquotesingle Aariz}, which is an Arabic word meaning \emph{cheek}. Since a cephalogram is an X-ray scan of the craniofacial area (i.e. cheek), we have taken \emph{\textquotesingle Aariz} in these terms. We believe that this dataset will not only derive forward research and innovation in automatic cephalometric landmark identification and CVM stage classification, but will also mark the beginning of a new era in the discipline.
    
    \section{Related Work}
    
    \subsection{Cephalometric Landmark Detection}
    With advances in machine learning and computer vision over the past few years, it has become increasingly clear that automatic detection of cephalometric landmarks could be a promising solution for improving inter-rater and intra-rater reliability, reducing tracing time, and enhancing orthodontic diagnosis and treatment planning. Recognizing the significance of automatic landmark detection, IEEE International Symposium on Biomedical Imaging (ISBI) hosted the challenge of Automatic Cephalometric Landmark Detection for Diagnosis in Cephalometric X-ray Images in both 2014 \cite{wang2014grand} and 2015 \cite{wang2015evaluation}. Wang et al. \cite{wang2016benchmark} introduced the first cephalometric landmark detection dataset\footnote{https://figshare.com/s/37ec464af8e81ae6ebbf}, comprising of 400 high-resolution X-ray images of patients ranging in age from six to 60 years. All images were obtained using the Soredex CRANEX Excel cephalometric X-ray machine with spatial dimensions of ${1935 \times 2400}$ pixels, at a resolution of 0.1 mm/pixel in both directions. Two orthodontists, with different levels of expertise, defined the ground truth annotations by reviewing the manual marking of 19 anatomical landmarks twice, and the mean of all four annotations was used as the ground truth, to account for inter-observer variability. However, there are several concerns regarding the generalizability and reliability of ground truth annotations. For instance, the training dataset only includes 150 cephalograms randomly selected from 400 patients with a wide range of ages (i.e. six to 60 years). Furthermore, all cephalograms are obtained from a single X-ray imaging device. With this limited amount of training data, an AI algorithm may struggle to generalize on such a diverse set of patients and be prone to overfit \cite{domingos2012few}. Additionally, the mean intra-observer variability of the senior and junior orthodontists is ${1.73 \pm 1.35}$ mm and ${0.90 \pm 0.89}$ mm, respectively, while the mean inter-observer variability between the two orthodontists is ${1.38 \pm 1.55}$ mm which produced a mean radial error (MRE) of ${2.02 \pm 1.53}$ mm on test data. This degree of variability is extremely large, considering the clinical precision range of $2$ mm. Consequently, there is a high probability that the trained model may contain unnecessary bias, suggesting that there is a limit to clinical applications merely with this dataset \cite{lee2020automated}. Despite these limitations, this dataset has served as a benchmark for the comparison of various cephalometric landmark detection models and laid the foundation of a new era in the field of dental X-ray imaging.

    In 2020, Zeng et al. \cite{zeng2021cascaded} published the PKU cephalogram dataset\footnote{https://doi.org/10.6084/m9.figshare.13265471.v1}, which consisted of 102 cephalograms from patients of ages 9 to 53 years. These images were captured using the Planmeca ProMax 3D cephalometric X-ray machine, with average spatial dimensions of ${2089 \times 1937}$ pixels and a resolution of approximately 0.125 mm/pixel. Two expert orthodontists independently annotated each image with 19 cephalometric landmarks. While this dataset was not particularly comprehensive, featuring only 102 cephalograms with the same 19 landmarks, it was still a valuable addition to the field of automated cephalometric landmark detection. 
    
    Aside from that, several other datasets containing large numbers of cephalograms annotated with an extensive amount of anatomical landmarks have also been offered and used on occasion. For example, Qian et al. \cite{qian2020cephann} proposed a multi-head attention neural network, CephaNN, for cephalometric landmark detection and used a private collection of 400 cephalograms with varying features and properties from various devices to evaluate the robustness of their proposed method. An experienced orthodontist manually labelled 75 landmarks on each image four times, and an average of all four annotations was used as the ground truth. While this dataset was annotated with almost all of the cephalometric landmarks and could be a valuable resource for the field of cephalometry, it may not have been made public due to concerns about patient privacy. Similarly, He et al. \cite{he2021cephalometric} demonstrated the significant translational invariance of their proposed architecture using a private dataset called the Huaxi-Analysis dataset. This dataset consisted of 1005 cephalometric images divided into training, validation and testing sets with 605, 200, and 200 images, respectively. Each image featured 37 landmarks and has spatial dimensions of ${1752 \times 1537}$ pixels with a pixel spacing of 0.13 mm. The dataset was labelled in two rounds by experienced orthodontists from the West China Hospital of Stomatology. In the first stage, the orthodontists manually located the landmarks in 512 radiographic images. In the second semi-automatic stage, a cephalometric landmark detection (CLD) model was trained using these 512 annotated radiographs to generate pre-located landmarks, which were then reviewed and adjusted by orthodontists. However, this dataset was also used only to evaluate the robustness of the proposed architecture and was not made public. Kunz et al. \cite{Kunz2019artificial} created a dataset with 1792 cephalometric X-ray images acquired from Sirona Orthophos XG radiographic machine. On each image, twelve examiners identified and marked a total of 18 radiographic landmarks. To verify inter- and intra-rater reliability, 20 cephalometric X-ray images were analyzed twice by each examiner to ensure high-quality training data. This dataset was the largest dataset ever annotated following excellent marking protocols, but unfortunately, it was not made available, and the research community was once again deprived of a sophisticated dataset. It is clear that, despite the creation of several datasets with extensive landmark annotations, they were specifically used for the evaluation of proposed models and have not been made available, leading to the continued lack of a comprehensive and reliable dataset.
    
    The story does not end here, as there is another very important point that requires discussion. In the field of orthodontics, the soft tissue paradigm \cite{almansob2019patient} has led to the consideration of facial soft tissue in all types of jaw and tooth movements. As a result, cephalometric studies include various soft tissue parameters, such as facial convexity, nasolabial angle, the position of upper and lower lips, mentolabial sulcus, position of soft tissue chin and lower anterior face height \cite{darkwah2018cephalometric} etc. These parameters are important for making orthodontic decisions about extraction and non-extraction treatment \cite{moon2021extraction}, the extent of retraction of anterior teeth, growth changes, and surgical movements of the maxilla and mandible. However, the publicly available datasets only include four soft tissue landmarks, which are insufficient for most soft tissue cephalometric analyses. Additionally, these datasets do not include important occlusal landmarks, which are essential for constructing the occlusal plane, an important factor in orthodontic diagnosis and treatment planning that can change during treatment. Hence, there is a great need for a new cephalometric landmark detection dataset that can address these limitations and assist researchers in developing algorithms that can contribute to better cephalometric decision-making.
    
    \subsection{CVM Stage Classification}
    The CVM stage assessment method has been widely employed to evaluate the growth stage in lateral cephalometric radiographs \cite{hassel1995skeletal, baccetti2002improved}. However, some studies have questioned the reliability and reproducibility of this method \cite{nestman2011cervical} due to the continuous nature of the skeletal maturation process and the difficulty in distinguishing between CVM stages for borderline subjects. Without a high level of technical knowledge and experience, it is challenging for clinicians to use the CVM stage method accurately. In recent years, the use of convolutional neural networks (CNNs) \cite{schwendicke2019convolutional} and deep learning techniques in the medical domain, especially in the field of dentistry, have gained popularity. A number of research studies have employed artificial intelligence (AI) to automate the assessment of CVM stages. In one such study, Rahimi et al. \cite{mohammad2022deep} implemented two transfer learning models based on ResNet-101, which were independently fine-tuned to determine CVM stage and pubertal growth spurt using lateral cephalograms. The study used 890 cephalometric radiographs acquired using the ProMax Dimax 3 Digital Pan/Ceph device. To address the issue of limited data, the researchers also included 400 lateral cephalograms from the IEEE International Symposium on Biomedical Imaging 2015 grand challenge \cite{wang2015evaluation}.
    
    Another study by Seo et al. \cite{seo2021comparison} aimed to evaluate and compare the performance of six advanced deep learning models based on CNNs for CVM classification in lateral cephalograms. The study included 600 images, with 100 images for each stage, labelled by a radiologist with over 10 years of experience, using Baccetti's method \cite{baccetti2005cervical}. The collected images had a size of ${1792 \times 2392}$ pixels, providing a clear visualization of the cervical vertebrae, including C2, C3, and C4. Obtaining high-resolution images of these CVM stages is quite a challenging task, making the dataset used in this study a particularly valuable resource. Zhou et al. \cite{zhou2021development} used a unique approach to classify CVM stages. They had a large sample size of 1080 cephalometric images manually labelled for 13 anatomic landmarks by an examiner on two separate occasions, with a three-month interval between the labelling sessions. The labelled images were then used to train and test a Detnet architecture, which was able to automatically label 2 reference landmarks and 13 anatomical landmarks on each image. The CVM stage was then determined based on linear and ratio measurements calculated from the labelled landmarks. This type of manual labelling had not been previously done and resulted in a dataset that could have been extremely valuable for training and evaluating machine learning models for CVM stage classification. However, the lack of access to the dataset limits its usefulness in advancing the field of CVM stage classification.

    Despite the numerous studies that have been conducted on automated CVM stage assessment using artificial intelligence, the field still lacks a standard benchmark dataset. The datasets that have been used in past studies are diverse and not publicly available, making it difficult to compare the results of different approaches. In order to move forward in this field and develop reliable and accurate methods for automatic CVM stage assessment, it is essential to have a standard benchmark dataset that can be used as a reference point. This would be a major step forward in the development of automated CVM stage assessment tools and their use in clinical practice.
    
    As it stands, the current cephalometric datasets available to researchers are limited in their scope and capabilities. Therefore, to make significant strides in the field of cephalometric analysis, a new and improved dataset is necessary. One that is comprehensive and diverse, with a larger number of cephalograms and more extensive annotations. Taking a cue, we proposed a new state-of-the-art dataset for cephalometric analysis, featuring 1000 cephalometric X-ray images acquired from 7 different imaging devices with varying resolutions. It boasts the most extensive collection of annotated soft tissue landmarks ever included in a publicly available dataset, as well as the first standard resource for CVM classification. Our team of clinical experts studiously annotated each image with 29 commonly used anatomical landmarks in two methodical phases, making it a valuable tool for researchers working to develop AI solutions for morphometric analysis. A thorough comparison of our dataset against other available and non-available datasets is presented in Table \ref{tab:datasets-comparison-table}, highlighting the unique features and superior characteristics of our dataset.
    
    \begin{figure}[ht!]
        \centering
        \includegraphics[width=16.6cm]{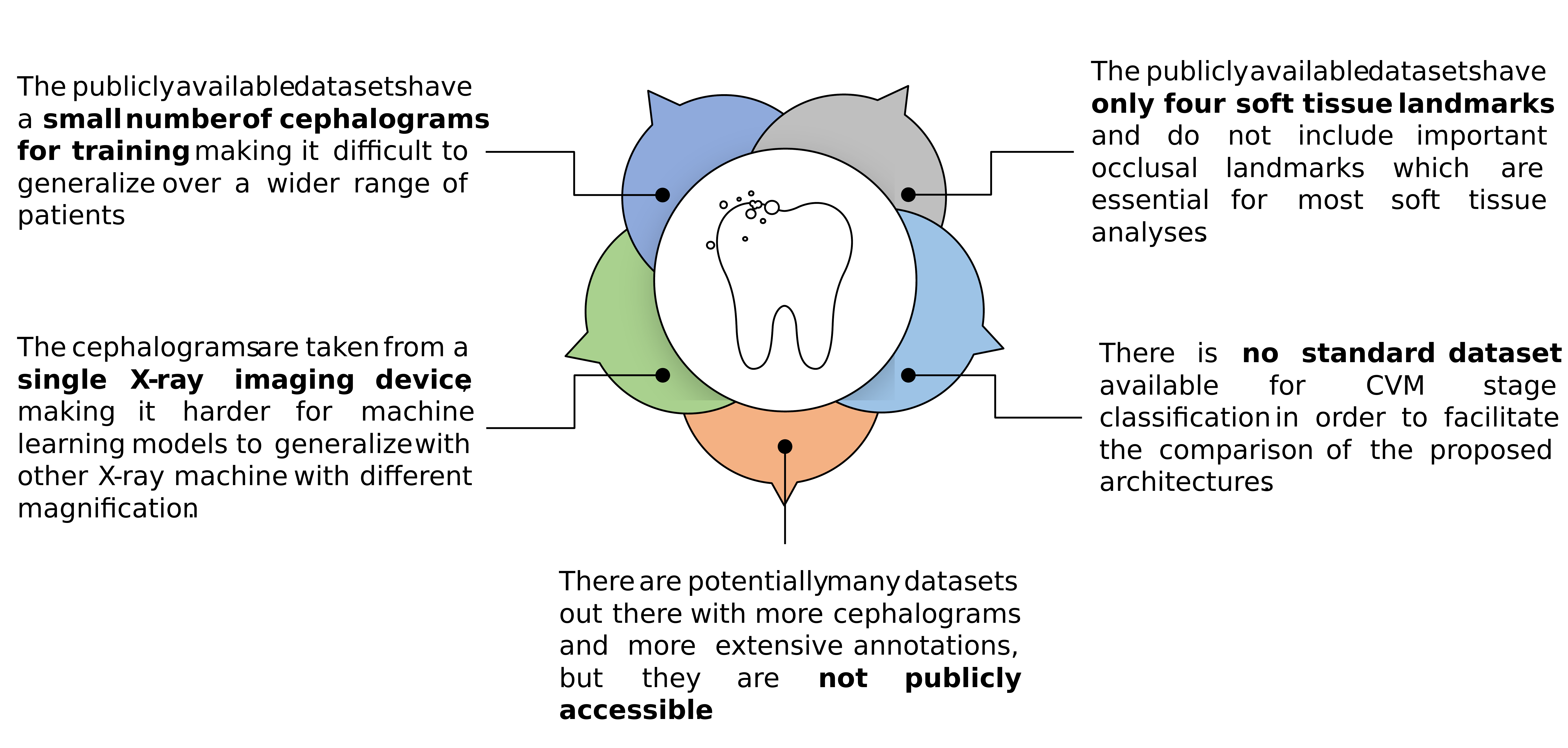}
        \caption{Summary of the literature review. The limitations reinforce the need for a new and improved dataset that can help researchers develop more accurate and reliable machine-learning models for cephalometric analysis.}
        \label{fig:literature-review-summary}
    \end{figure}
    
    \begin{table}[ht!]
        \centering
        \caption{A side-by-side comparison of cephalometric datasets. Our dataset stands out with its diversity, featuring images acquired from various X-ray machines and annotations for both anatomical landmark detection and CVM stage classification.}
        \begin{tabular}{lccccc}
            \toprule
            Datasets & Cephalograms & Landmarks & X-ray imaging devices & Available & CVM stage method \\
            \midrule
            Wang et al. \cite{wang2016benchmark}    & 400           & 19            & 1             & \cmark            & \xmark            \\
            Zeng et al. \cite{zeng2021cascaded}     & 102           & 19            & 1             & \cmark            & \xmark            \\
            Qian et al. \cite{qian2020cephann}      & 400           & 75            & N/A           & \xmark            & \xmark            \\
            He et al. \cite{he2021cephalometric}    & 1005          & 37            & N/A           & \xmark            & \xmark            \\
            Kunz et al. \cite{Kunz2019artificial}   & 1792          & 18            & 1             & \xmark            & \xmark            \\
            \midrule
            \textbf{\emph{\textquotesingle Aariz}}  & \textbf{1000} & \textbf{29}   & \textbf{7}    & \textbf{\cmark}   & \textbf{\cmark}   \\
            \bottomrule
      \end{tabular}
      \label{tab:datasets-comparison-table}
    \end{table}

    \section{Dataset Description}
    The scarcity of high-quality cephalometric datasets for research purposes is primarily because of the difficulty in obtaining access to medical images due to patients' privacy concerns. Even if one is successful in obtaining access to these images, the process of annotating them can be a costly and time-consuming endeavour. However, the significance of such datasets cannot be understated, as they provide valuable resources for researchers seeking to enhance automated AI systems used in cephalometric analysis. In an effort to address these problems, we embarked on a research collaboration with Riphah International University in 2020. The main objective of this research was to create a benchmark dataset that not only addresses the limitations of existing datasets but also helps improve the performance of cephalometric landmark tracing algorithms. An ethical approval was received from the Institutional Review Committee of Islamic International Dental College, Riphah International University, Islamabad, Pakistan (IRB Number IIDC/IRC/2020/001/012). 

    We have proposed a new benchmark dataset consisting of 1000 cephalometric X-ray images, which were collected from 1000 patients ranging in age from 8 to 62 years, and acquired from 7 different X-ray imaging devices with varying resolutions. Table \ref{tab:xray-imaging-devices} summarizes all of the X-ray imaging devices used to obtain the cephalograms, as well as their respective resolutions and the number of cephalograms from each machine. Each cephalogram is labelled with 29 cephalometric landmarks, which are most commonly used in clinical settings and can be employed for a variety of cephalometric measurements. These landmarks are selected from 3 fundamental categories of anatomical structures: skeletal, dental and soft-tissue. The skeletal structure contributes 15 landmarks, whereas 8 landmarks are related to dental structures, and 6 landmarks are related to soft-tissue structures, making it the most comprehensive dataset with the highest number of soft-tissue landmarks in any publicly available data resource to date. In Table \ref{tab:cephalometric-landmarks-descriptions}, we have compiled a list of the cephalometric landmarks that are included in our dataset, organized by their respective categories, and accompanied by their clinical definitions to provide a clear understanding of their role in cephalometric analysis. 

    \begin{table}[!th]
        \centering
        \caption{A breakdown of the diversity in our cephalometric dataset, featuring the distribution of images acquired from various X-ray machines along with their manufacturers and resolutions.}
        \begin{tabular}{ccccc}
            \toprule
            No. & Machine & Manufacturer & Resolution (millimetres/pixel)  & Cephalograms \\
            \midrule
            1. & ART Plus               & BLUEX     & 0.1   & 366   \\
            2. & Veraviewepocs 2D       & J. Morita & 0.144 & 177   \\
            3. & Smart3D                & LargeV    & 0.1   & 59    \\
            4. & ProMax 2D              & Planmeca  & 1.139 & 41    \\
            5. & ProMax with ProTouch   & Planmeca  & 0.139 & 135   \\
            6. & Hyperion X5            & Myray     & 0.089 & 143   \\
            7. & Rotograph EVO          & Villa     & 0.135 & 79    \\
            \bottomrule
        \end{tabular}
        \label{tab:xray-imaging-devices}
    \end{table}

    \begin{figure}[!ht]
        \centering
        \begin{tabular}{cc}
            \subfloat[]{\includegraphics[ width=9cm]{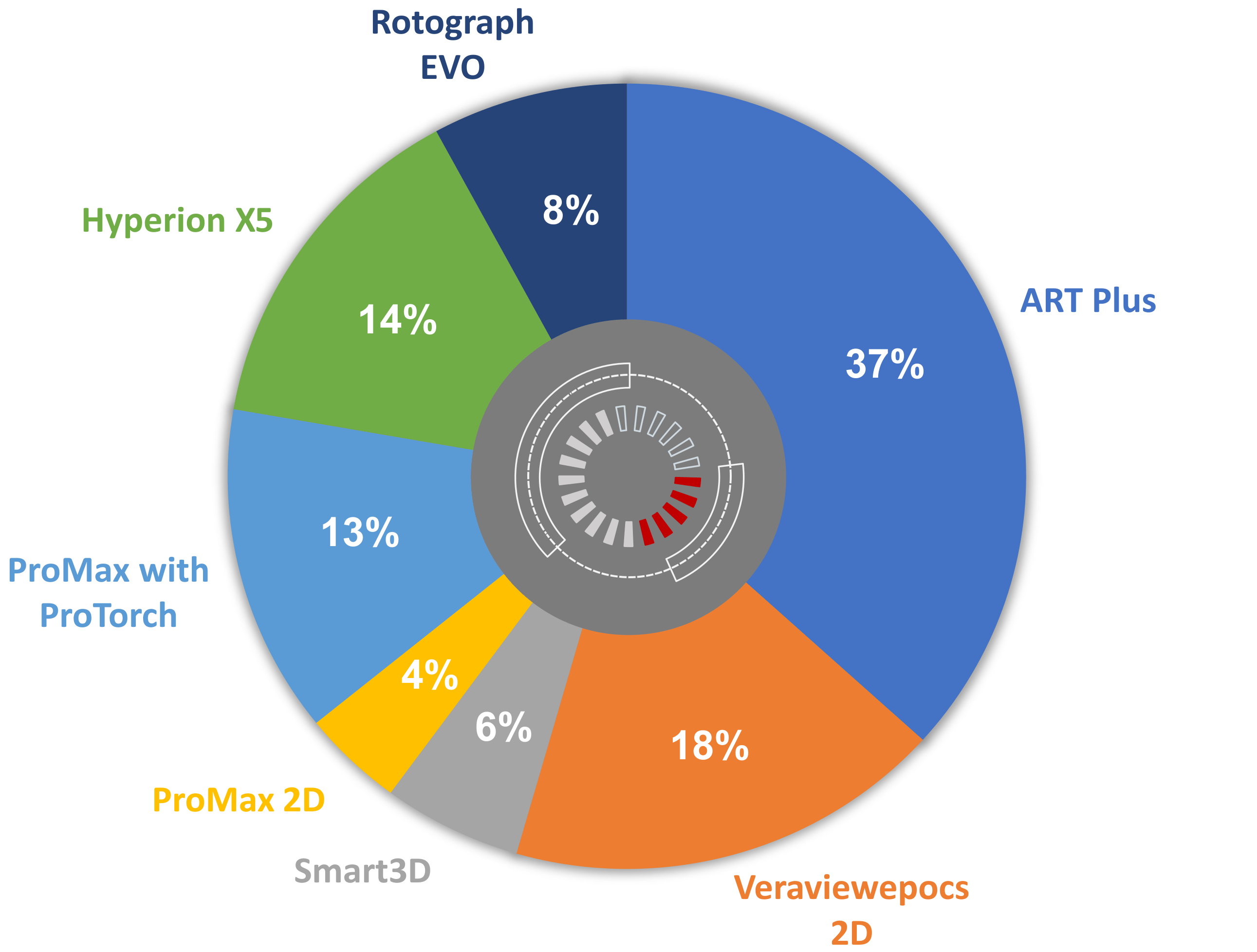}} &
            \subfloat[]{\includegraphics[width=6.8cm]{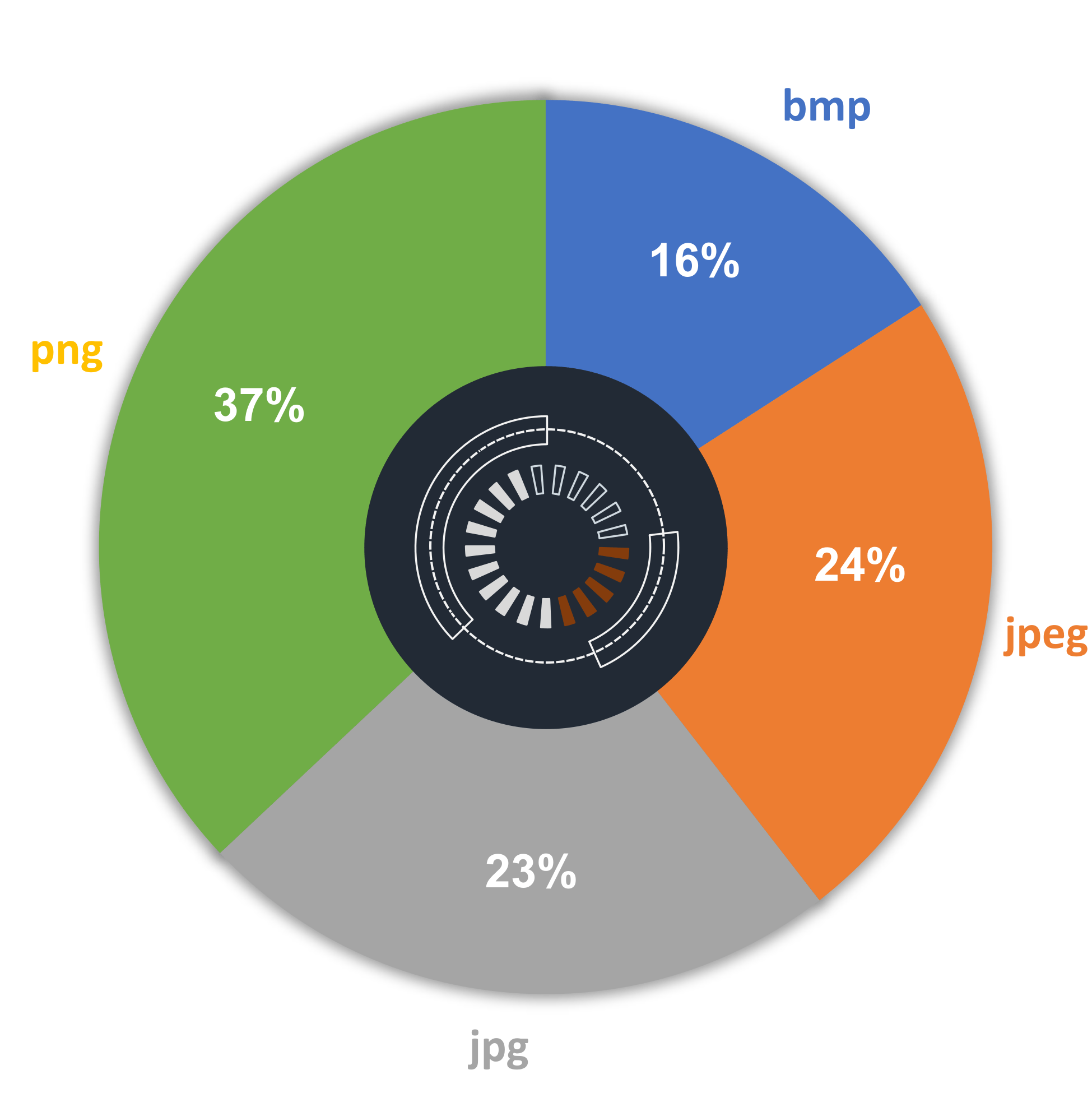}}
        \end{tabular}
        \caption{A closer look at the distribution of source and format of images in our cephalometric dataset. (a) illustrates the proportions of images obtained from each cephalometric X-ray machine. (b) shows the distribution of image formats in the dataset.}
        \label{fig:dataset-descriptive-features}
    \end{figure}

    To ensure the accuracy and consistency of dataset annotations, we enlisted the help of a team of skilled clinicians from the Islamic International Dental College in Islamabad, Pakistan. Comprised of 2 expert orthodontists with a combined clinical experience of 25 years, as well as 4 additional orthodontic professionals, this team was responsible for deliberately annotating each of the 1000 cephalograms in the dataset. The expert orthodontists also presided over the entire annotation process, occasionally participating in the labelling themselves to ensure the highest level of quality and consistency. The overall dataset is  divided into three subsets: training, validation and testing, with 700, 150, and 150 cephalometric X-ray radiographs, respectively. The images from each of the X-ray machines have different spatial dimensions and are uniformly distributed in each subset. This ensures that our dataset is well-rounded and representative of the diverse range of images collected from various X-ray machines.

    \begin{table}[!ht]
        \centering
        \caption{A list of annotated cephalometric landmarks in our dataset, including detailed clinical descriptions.}
        \begin{tabular}{p{0.5cm}p{3.5cm}p{1cm}p{2cm}p{7.5cm}}
            \toprule
            No. & Landmarks & Symbol & Category & Clinical Description \\
            \midrule
            1.  & A-point (Subspinale)      & A     & Skeletal      & The innermost point on the contour of the pre-maxilla between the anterior nasal spine and the incisor tooth.                         \\
            2.  & Anterior Nasal Spine      & ANS   & Skeletal      & Most anterior point of the osseous anterior nasal spine in the median-sagittal plane i.e. furthest anterior point of the maxilla. \\
            3.  & B-point (Supramentale)    & B     & Skeletal      & The innermost point on the contour of the mandible between the incisor tooth and the bony chin.                                       \\
            4.  & Menton                    & Me    & Skeletal      & The most inferior point on the mandibular symphysis-that is the bottom of the chin.                                                   \\
            5.  & Nasion                    & N     & Skeletal      & The anterior point of the intersection between the nasal and frontal bones.                                                           \\
            6.  & Orbitale                  & Or    & Skeletal      & The lowest point on the inferior margin of the orbit.                                                                                 \\
            7.  & Pogonion                  & Pog   & Skeletal      & The most anterior point on the contour of the chin.                                                                                   \\
            8.  & Posterior Nasal Spine     & PNS   & Skeletal      & The tip of the posterior spine of the palatine bone, at the junction of the hard and soft palates.                                    \\
            9.  & Ramus                     & R     & Skeletal      & The most convex point on the exterior border of the ramus along the vertical.                                                         \\
            10. & Sella                     & S     & Skeletal      & The midpoint of the cavity of sella turcica.                                                                                          \\
            11. & Articulare                & Ar    & Skeletal      & The point of intersection between the shadow of the zygomatic arch and the posterior border of the mandibular ramus.                  \\
            12. & Condylion                 & Co    & Skeletal      & Most posterior/superior point on the condyle of the mandible.                                                                         \\
            13. & Gnathion                  & Gn    & Skeletal      & Point located perpendicular on the mandibular symphysis midway between pogonion and menton.                                           \\
            14. & Gonion                    & Go    & Skeletal      & The midpoint of the contour connecting the ramus and body of the mandible.                                                            \\
            15. & Porion                    & Po    & Skeletal      & The midpoint of the upper contour of the external auditory canal (anatomic porion).                                                   \\
            16. & Lower 2nd PM Cusp Tip     & LPM   & Dental        & Buccal cusp tip of lower 2nd premolar.                                                                                                \\
            17. & Lower Incisor Tip         & LIT   & Dental        & Insical edge of the lower central incisors.                                                                                           \\
            18. & Lower Molar Cusp Tip      & LMT   & Dental        & Mesio-buccal cusp tip of lower 1st molar.                                                                                             \\
            19. & Upper 2nd PM Cusp Tip     & UPM   & Dental        & Buccal cusp tip of upper 2nd premolar.                                                                                                \\
            20. & Upper Incisor Apex        & UIA   & Dental        & Apical root tip of upper central incisors.                                                                                            \\
            21. & Upper Incisor Tip         & UIT   & Dental        & Insical edge of the upper central incisors.                                                                                           \\
            22. & Upper Molar Cusp Tip      & UMT   & Dental        & Mesio-buccal cusp tip of upper 1st molar.                                                                                             \\
            23. & Lower Incisor Apex        & LIA   & Dental        & Apical root tip of lower central incisors.                                                                                            \\
            24. & Labrale inferius          & Li    & Soft Tissue   & Most prominent point on the vermilion border of the lower lip in midsagittal plane.                                                           \\
            25. & Labrale superius          & Ls    & Soft Tissue   & Most prominent point on the vermilion border of the upper lip in the midsagittal plane.                                                           \\
            26. & Soft Tissue Nasion        & N`    & Soft Tissue   & Point on soft tissue over nasion.                                                                                                     \\
            27. & Soft Tissue Pogonion      & Pog`  & Soft Tissue   & Soft tissue over pogonion.                                                                                                            \\
            28. & Subnasale                 & Sn    & Soft Tissue   & In the midline, the junction where the base of the columella of the nose meets the upper lip.                                             \\
            29. & Pronasale                 & Pn    & Soft Tissue   & Tip or apex of external nose. \\
            \bottomrule
        \end{tabular}
        \label{tab:cephalometric-landmarks-descriptions}
    \end{table}

    \begin{figure}[!ht]
        \centering
        \begin{tabular}{cc}
            \subfloat[]{\includegraphics[width=7.1cm]{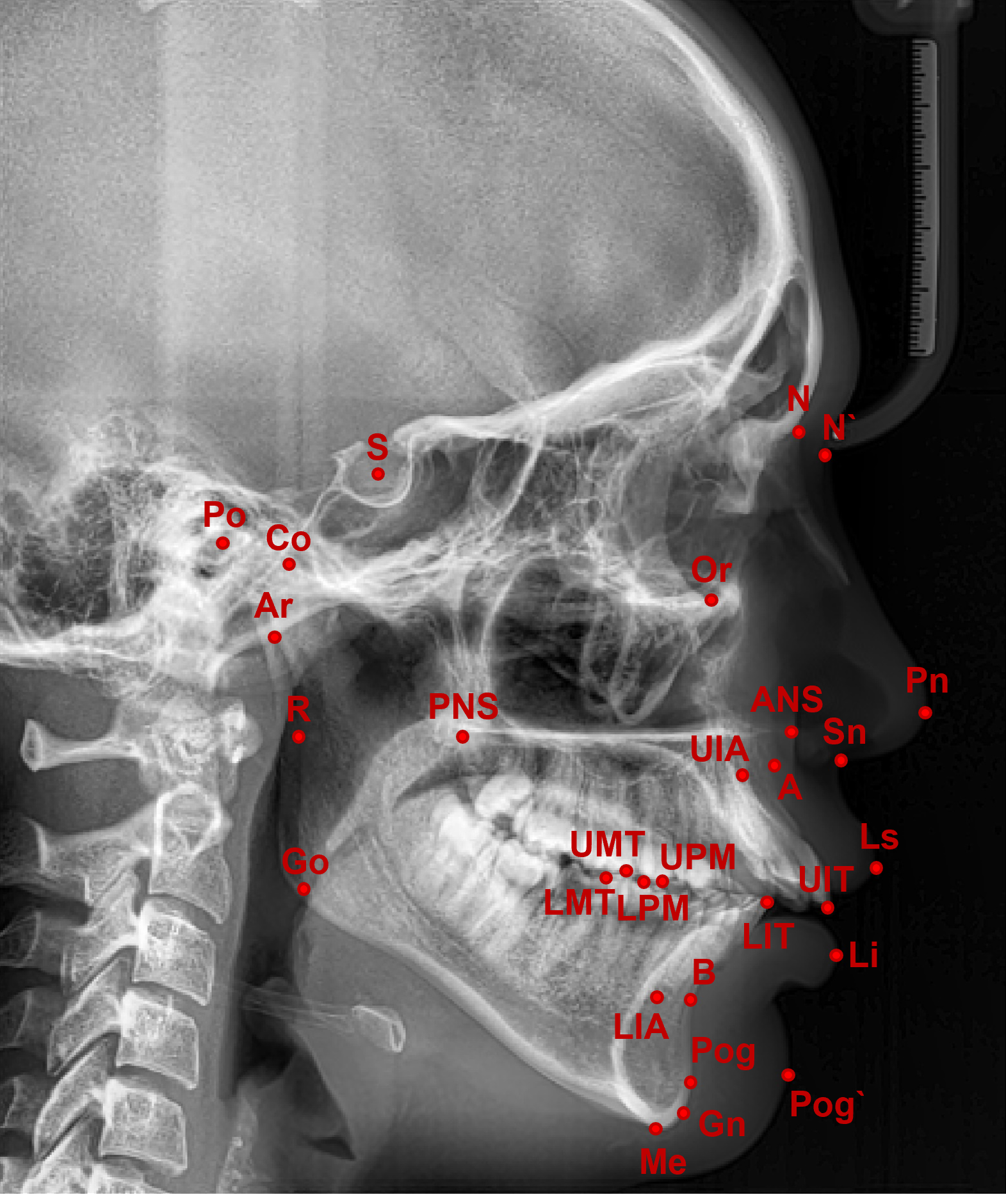}} &
            \subfloat[]{\includegraphics[width=7.8cm]{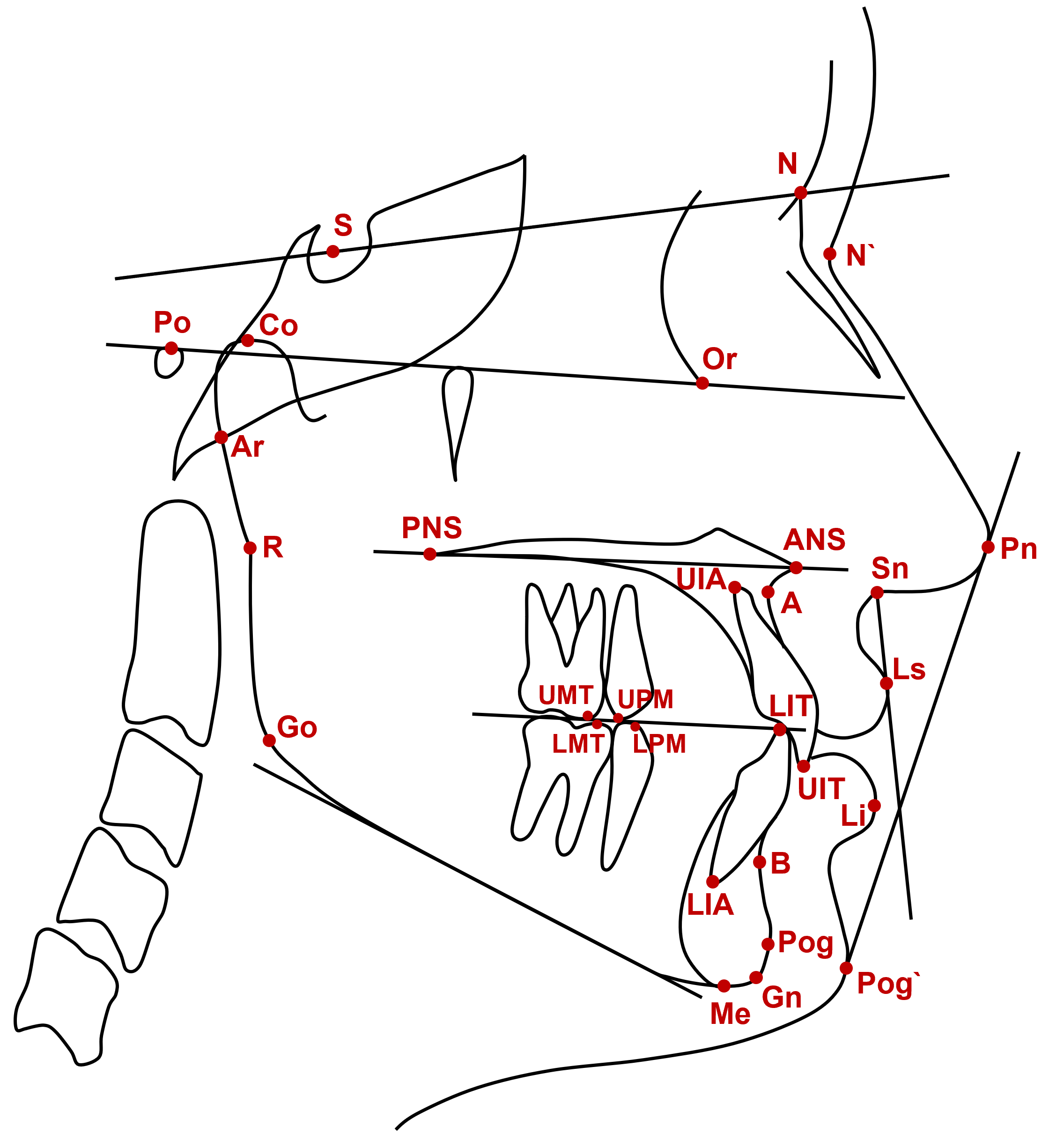}}
        \end{tabular}
        \caption{A visualization of anatomical landmarks and cephalometric tracing (a) showcases a cephalogram from our dataset with all 29 landmarks. (b) demonstrates the tracing of various cephalometric structures in a sample image.}
        \label{fig:annotated-traced-cephalogram}
    \end{figure}

    \subsection{Data Acquisition}
    The dataset consists of lateral cephalometric X-ray images collected from the archives of patients undergoing orthodontic treatment at Islamic International Dental College, Islamabad, Pakistan. We obtained written consent from all patients for the use of their radiographic records for educational and research purposes, and ensured the confidentiality of their records by obscuring identifying information. Table \ref{tab:inclusion-exclusion-criteria} features the inclusion and exclusion criteria that we followed during data collection.

    \begin{table}[!ht]
        \centering
        \caption{The inclusion and exclusion criteria for selecting cephalometric radiographs to ensure high quality and relevance in our data collection process.}
        \begin{tabular}{p{7cm}p{7cm}}
            \toprule
            \multicolumn{1}{c}{Inclusion criteria} & \multicolumn{1}{c}{Exclusion criteria} \\
            \midrule
            Only those images that met the following criteria were included in the dataset:
            \begin{itemize}
                \item Radiographs adequately showing the area of interest i.e. skeletal, dental, and soft tissue structures.
                \item Radiographs having all permanent teeth erupted till the first permanent molars in the maxillary and mandibular arch.
                \item Radiographs without any facial cleft, syndrome or dentofacial deformity
            \end{itemize} & 
            The images that had the following characteristics were excluded from the dataset:
            \begin{itemize}
                \item Radiographs with congenitally missing, extracted, impacted and supernumerary teeth.
                \item Radiographs that had heavily restored teeth with restoration, crowns or veneers involving cusp tips of first molars.
                \item Radiographs of patients with mixed dentition stage.
                \item Radiographs not showing cervical spine up to the fourth cervical vertebrae.
                \item Radiographs with artefacts obscuring the interpretation of the image.
            \end{itemize} \\
        \bottomrule
        \end{tabular}
        \label{tab:inclusion-exclusion-criteria}
    \end{table}

    Among 3500 cephalometric radiographs, our junior orthodontists curated 1000 radiographs that fulfil the inclusion and exclusion criteria. We matched the radiographs with their corresponding imaging machines and assigned them anonymous identifiers to protect patient privacy. The final dataset was then expertly reviewed and approved by senior orthodontists, marking the beginning of the labelling phase.
    
    \subsection{Dataset Annotation Process}
    Before diving into the annotation process, our team of expert orthodontists conducted a rigorous training session, during which they emphasized the importance of adhering to standard cephalometric terminology and highlighted the most commonly confused landmarks. All the clinicians were required to annotate a set of preliminary test images, which were then reviewed by expert orthodontists. Any annotations that did not meet the required standards were corrected and the clinicians received additional feedback and guidance. This process was repeated until all clinicians achieved the desired proficiency level in confidently distinguishing between commonly confused anatomical structures. With such a solid foundation in proper cephalometric annotation techniques and protocols, our team of clinicians was ready to embark on the meticulous task of annotating the real dataset, which culminated in a series of reviews and corrections to ensure the highest level of accuracy.
        
        \subsubsection{Cephalometric Landmarks}
\begin{figure}[!ht]
            \centering
            \includegraphics[width=13cm]{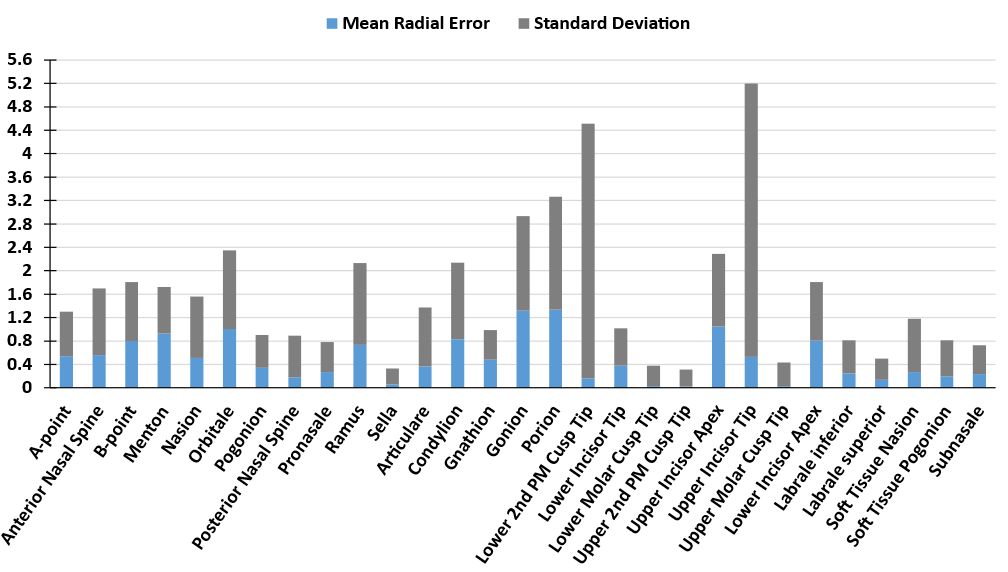}
            \caption{Uncovering the variability: A look at the inter-observer differences between junior and senior orthodontists through the lens of mean radial error and standard deviation.}
            \label{fig:labellers-vs-reviewers}
        \end{figure}

        The annotation process of cephalometric landmarks was carried out in two phases. In the first phase, two junior orthodontists independently marked all of the cephalograms, and in the second phase, two senior orthodontists collaboratively reviewed and corrected the markings as needed. The average of the markings by junior orthodontists and the average of the markings by senior orthodontists are provided separately. The mean of these two averaged markings will be used as the ground truth for cephalometric landmarks. The team utilized LabelBox\footnote{https://labelbox.com/} as a labelling tool to annotate the cephalograms. At the end of the annotation process, inter-observer variability between junior and senior orthodontists was found to be ${0.494 \pm 1.097}$ mm in terms of mean radial error $\pm$ standard deviation. Figure \ref{fig:labellers-vs-reviewers} features the landmark-wise distribution of inter-observer variabilities between the two groups of clinicians.

        \subsubsection*{Labelling Phase:}
\begin{figure}[!t]
            \centering
            \begin{tabular}{cc}
                \hspace*{-1.5cm}\includegraphics[width=9.3cm]{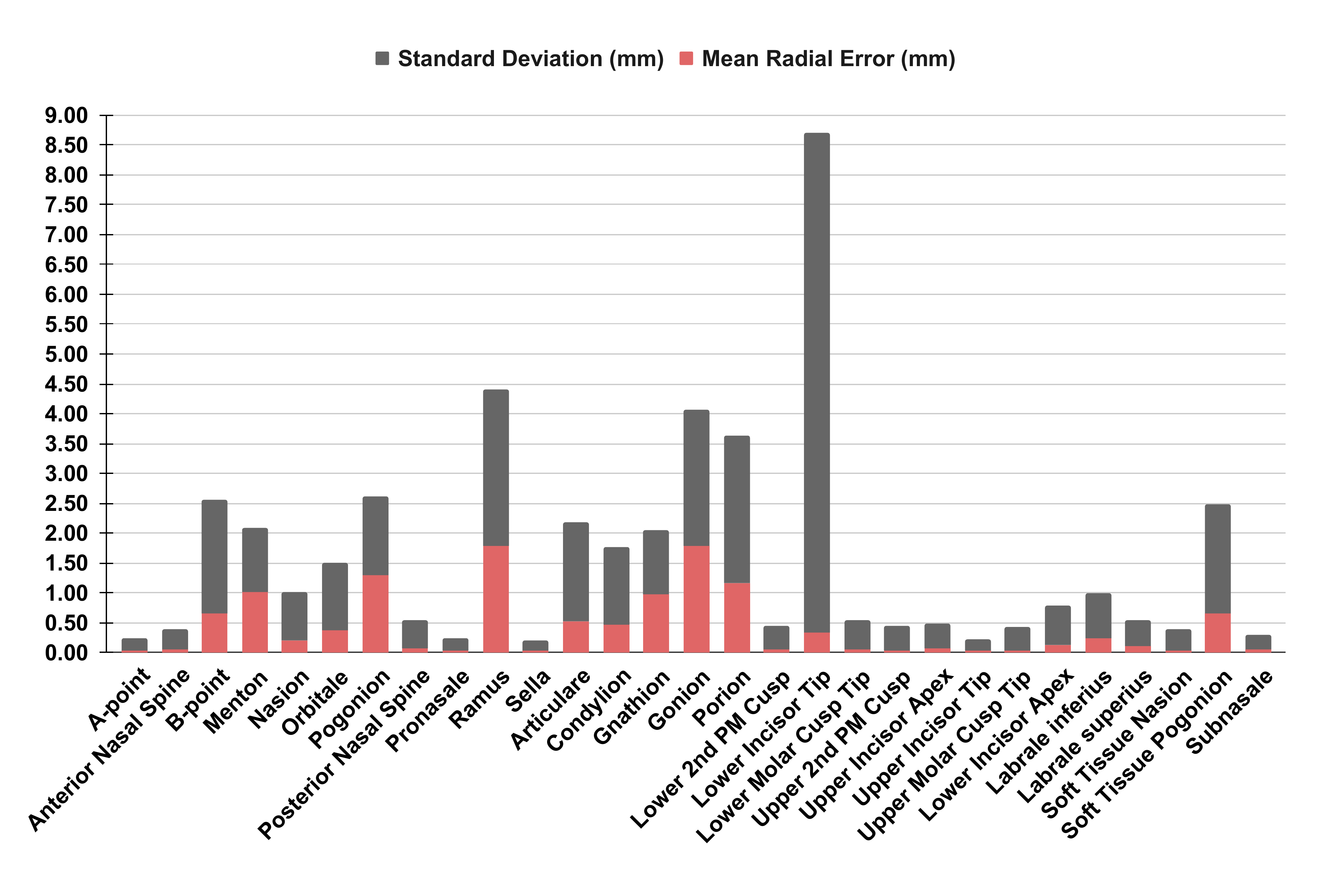} & \includegraphics[width=9.3cm]{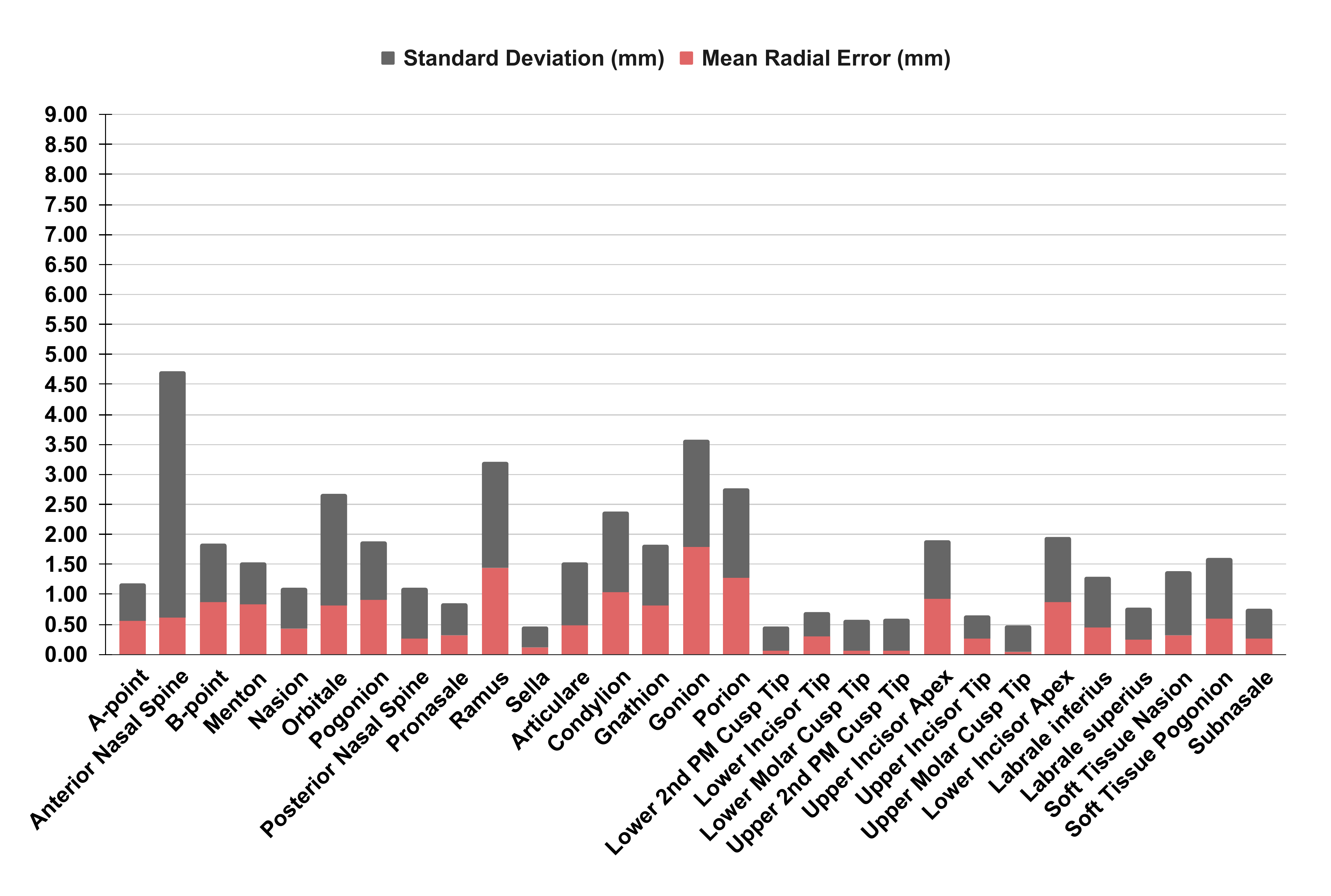} \\
                \hspace*{-1.5cm}\includegraphics[width=9.3cm]{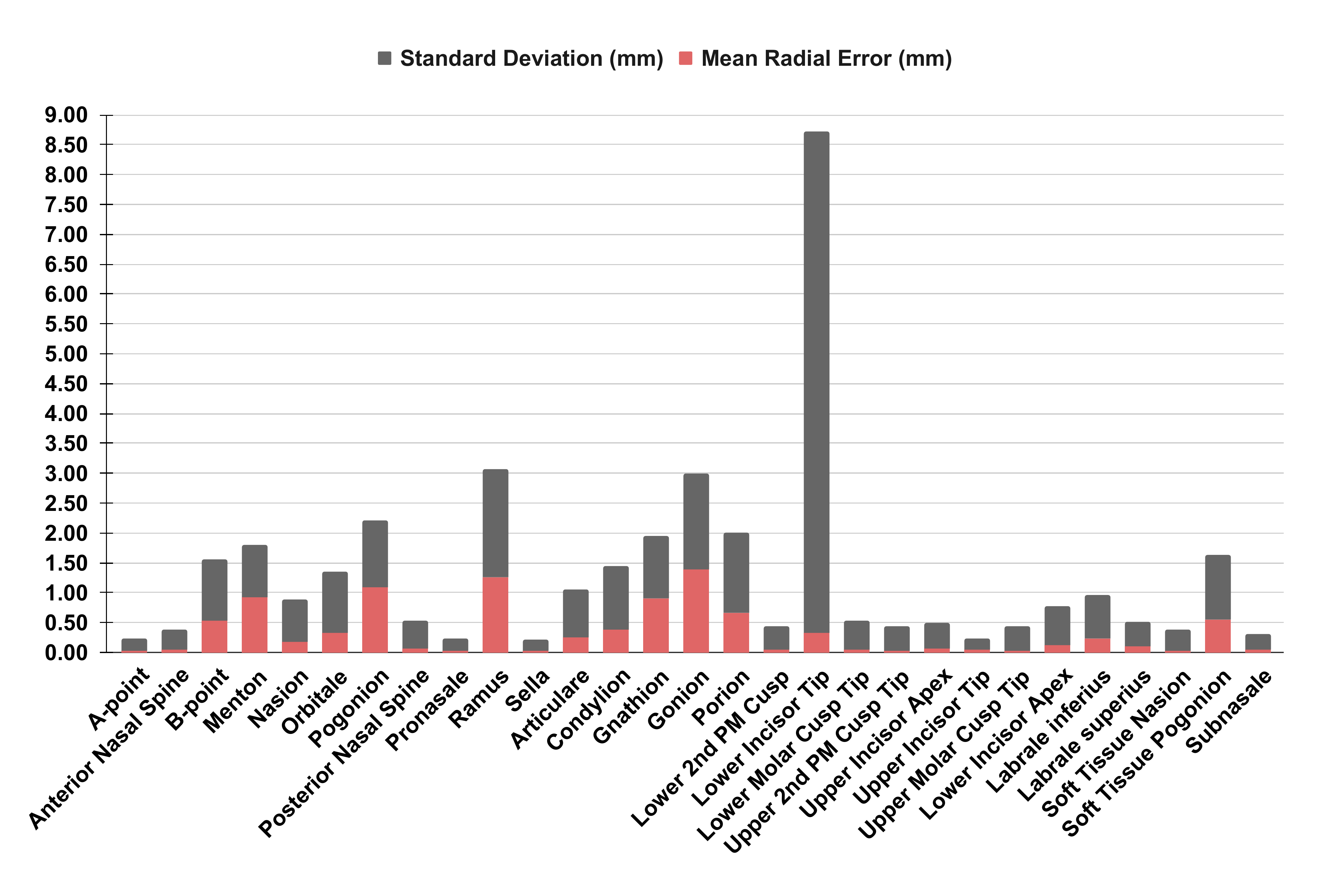} & \includegraphics[width=9.3cm]{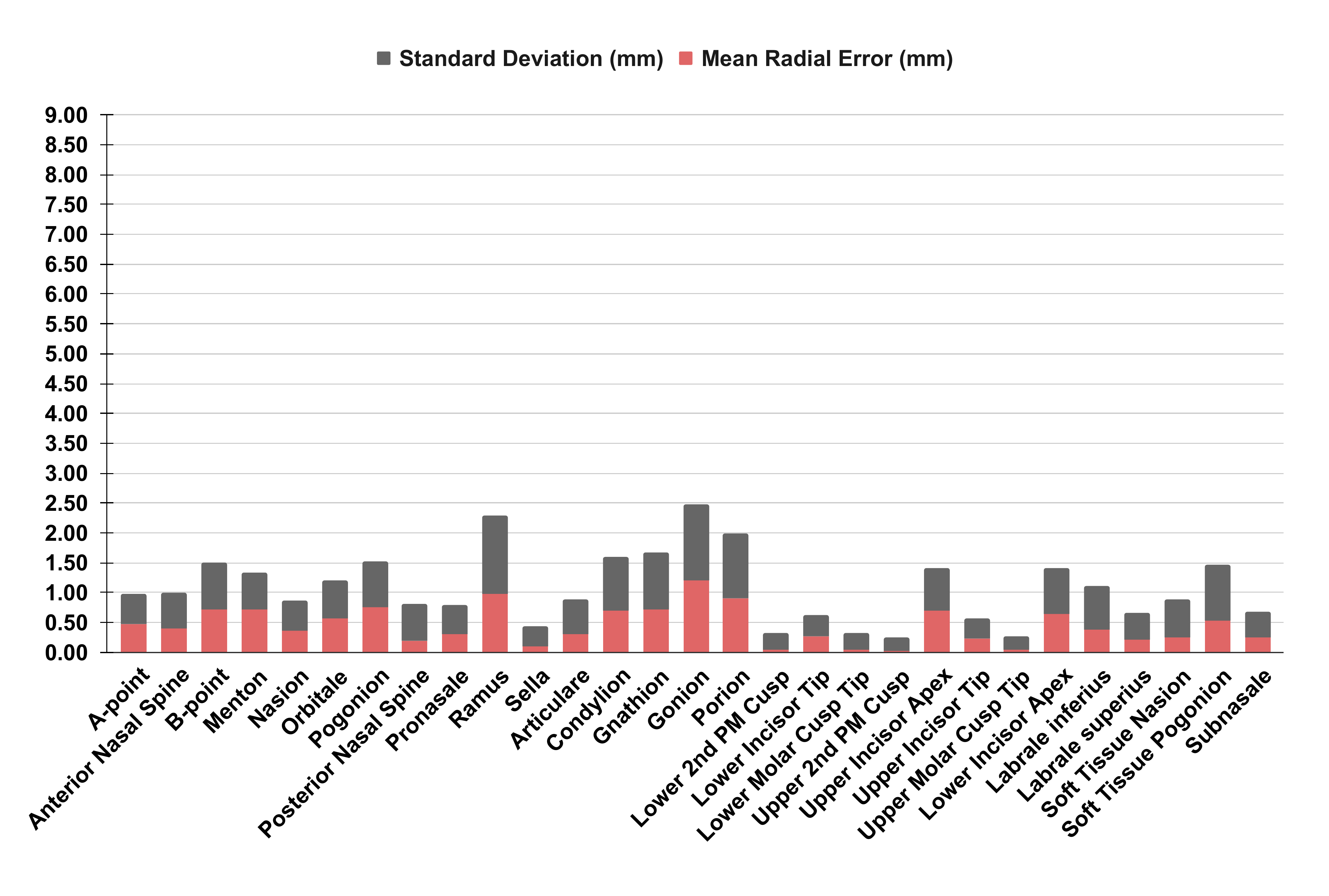} \\
                \hspace*{-1.5cm}\subfloat[]{\includegraphics[width=9.3cm]{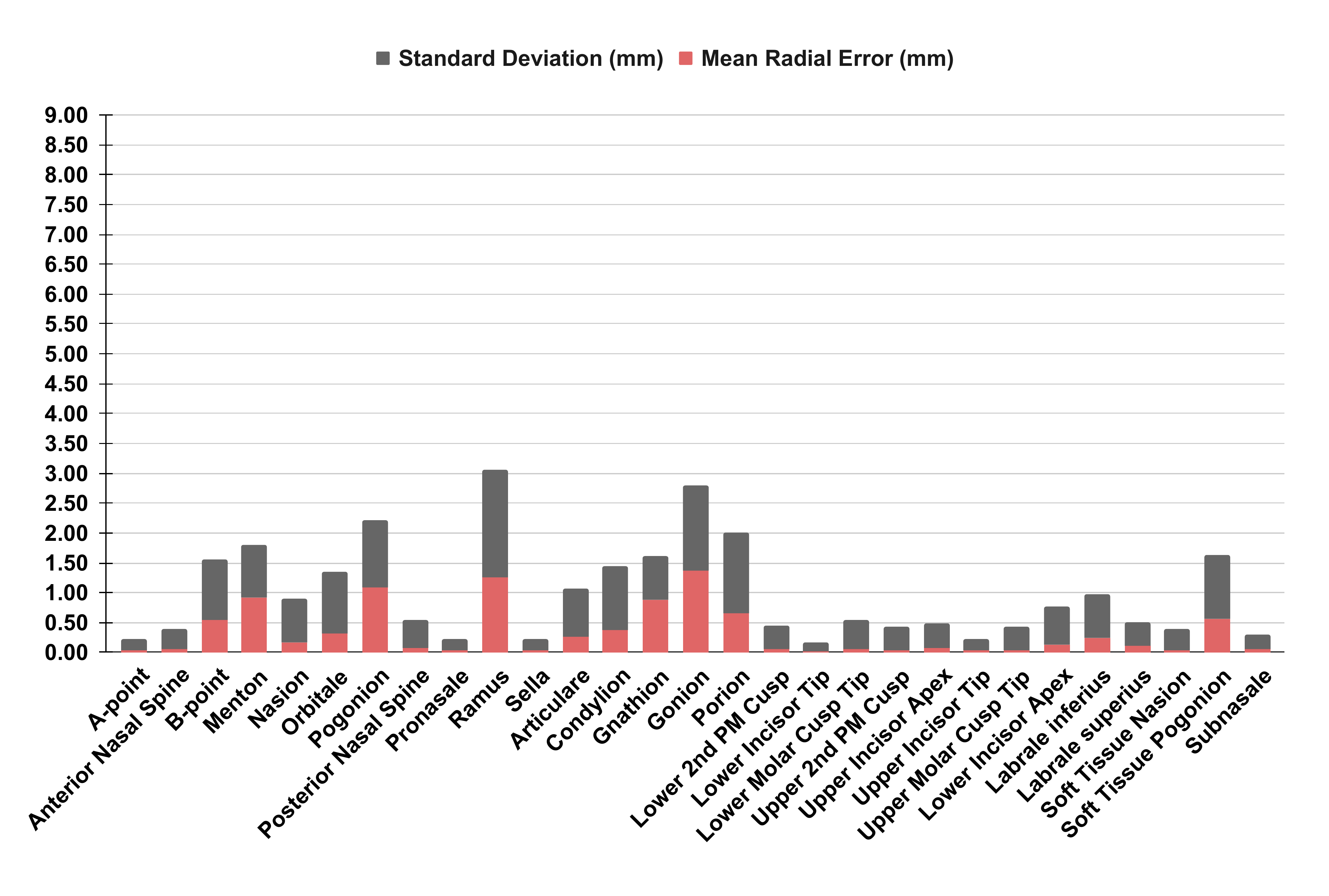}} & \subfloat[]{\includegraphics[width=9.3cm]{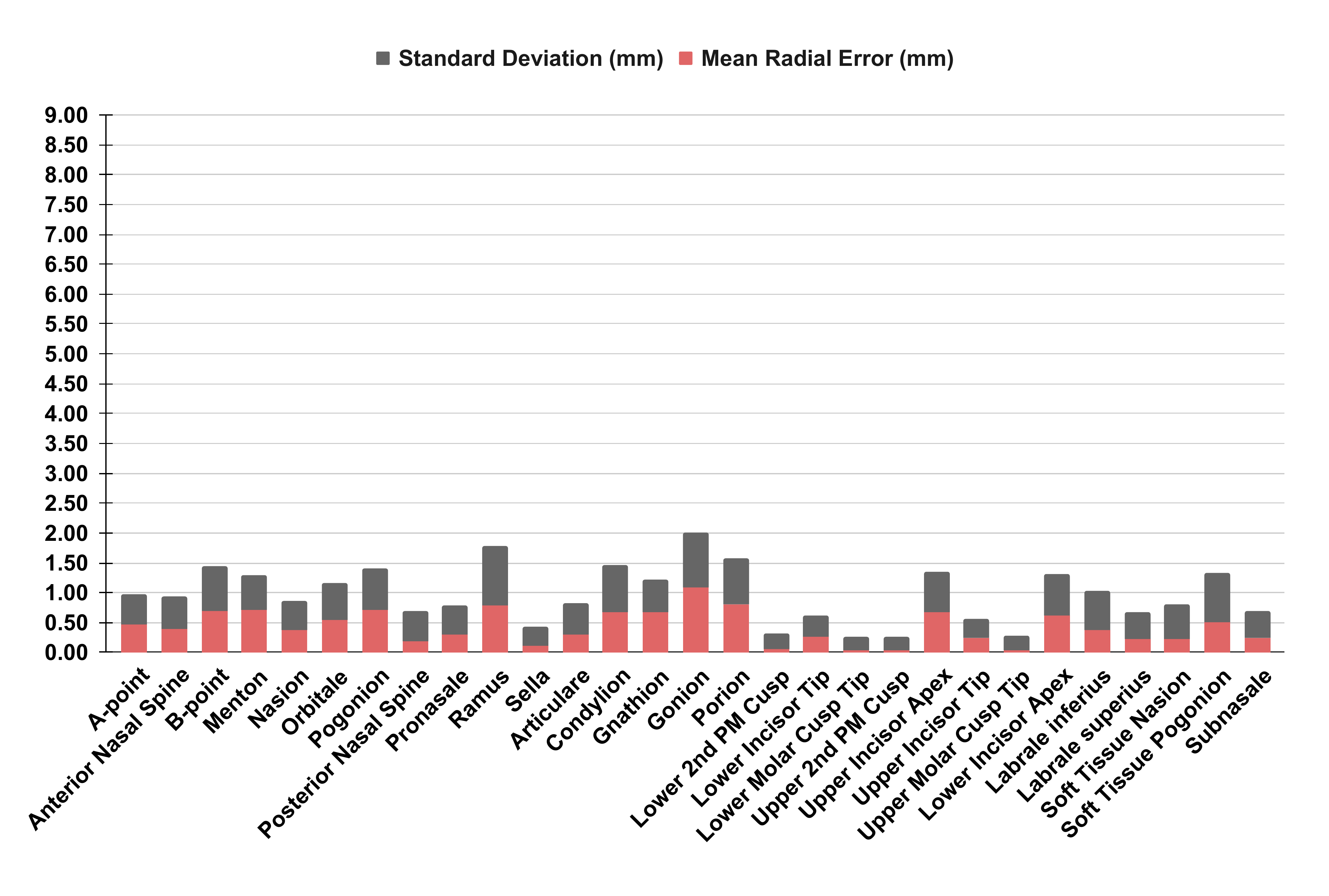}}
            \end{tabular}
            \caption{A visual representation of inter-observer variability in landmark annotations. The graphs showcase the precision of our annotation process across all rounds of the labelling and reviewing phases. \textbf{(left)}  The top, middle, and bottom rows show the variability among junior orthodontists during the three rounds of the labelling phase. \textbf{(right)} The top, middle, and bottom rows depict the variability among senior orthodontists during all the rounds of the reviewing phase.}
            \label{fig:lebelling-reviewing-phases-variability-graphs}
        \end{figure}

         The first phase of cephalometric landmark annotations, i.e. labelling phase, was accomplished in three rounds. Figure \ref{fig:lebelling-reviewing-phases-variability-graphs} (a) illustrates the state of inter-observer variability with respect to each landmark during all these rounds. To begin with labelling, two junior orthodontists with five years of clinical experience, referred to as labellers, independently traced each cephalogram with 29 anatomical landmarks. This first round of labelling was statistically reviewed and the inter-observer variability was determined to be $0.425 \pm 1.170$ mm in terms of mean radial error (MRE) $\pm$ standard deviation (SD). However, as shown in Figure \ref{fig:lebelling-reviewing-phases-variability-graphs} (a) (top), it was observed that the inter-observer variability between the labellers in terms of mean radial error was quite high for some landmarks (e.g. Ramus, Gonion and Porion etc). Therefore, to further improve markings, a total of 601 cephalograms were identified for which the mean radial error of critical landmarks was found to be greater than a threshold (i.e. 50 pixels or approximately 5 mm). The erroneous landmarks were re-traced by labellers under the supervision of expert orthodontists. As a result, mean inter-observer variability was significantly reduced to $0.341 \pm 0.964$ mm. Finally, we focused on reducing the standard deviation between landmark annotations. Based on statistical analysis, it was revealed that the standard deviation for the lower incisor tip was significantly large throughout the labelling phase, leading to the identification of 12 cephalograms with incorrect markings for this landmark. These were subsequently corrected by expert orthodontists, resulting in a significant reduction in standard deviation for that landmark. Through this multi-stage labelling process, we aimed to reduce inter-observer variability and ensure the accuracy and reliability of the labelling phase.
    
         To calculate the intra-observer variabilities between the orthodontists, we randomly selected 100 cephalometric images from the dataset, one image from every 10 images, as a representative sample. We then asked the labellers to re-annotate the images in this subset, assuming that it reflects the characteristics of the entire dataset. As a result, the intra-observer variability was estimated by comparing these new annotations to their corresponding previous ones. Table \ref{tab:labelling-phase-variabilities} presents the mean intra- and inter-observer variabilities of both labellers at the end of the labelling phase.
    
        \begin{table}[!th]
            \centering
            \caption{Mean intra- and inter-observer variabilities of labellers in terms of Mean Radial Error (MRE) $\pm$ Standard Deviation (SD) at the end of the labelling phase.}
            \begin{tabular}{cccc}
                \toprule
                {} & \multicolumn{2}{c}{Intra-observer Variability} & Inter-observer Variability \\ 
                \cmidrule(r){2-4}
                {} & Labeller 1 & Labeller 2 & Labeller 1 vs. Labeller 2 \\ 
                \midrule
                MRE (mm)  & {1.473 $\pm$ 1.829} & {1.651 $\pm$ 2.003} & {$0.329 \pm 0.663$} \\ 
                \bottomrule
            \end{tabular}
            \label{tab:labelling-phase-variabilities}
        \end{table}

        \subsubsection*{Reviewing Phase:}
        The second phase of cephalometric landmark annotations, i.e. reviewing phase, was also carried out in three rounds. In this phase, two senior orthodontists with clinical expertise ranging from 7 to 10 years, referred to as reviewers, reviewed the annotations by labellers at the end of the labelling phase and corrected them as needed. Figure \ref{fig:lebelling-reviewing-phases-variability-graphs} (b) illustrates the trend of inter-observer variability with respect to each landmark during all these rounds. As described in the previous section, both labellers had annotated each cephalogram separately, therefore, each reviewer was required to review a total of 2000 labelled cephalograms, which was a daunting task. Therefore, the reviewers worked together, with one reviewing the markings of labeller 1 for the first 500 cephalograms and the markings of the other labeller for the remaining 500 cephalograms, while the other reviewer did the opposite. Through this collaborative approach, the reviewers were able to effectively review all of the cephalograms and made necessary corrections. The reviewing phase followed a similar pattern as the labelling phase, starting with a general review of the entire dataset, followed by the rounds focusing on reducing radial error and targeting the reduction of standard deviation. To measure the intra-observer variability of reviewers, we used the same subset of 100 cephalograms that had already been annotated by the labellers. The reviewers then collaboratively reviewed and assessed these annotations, allowing us to calculate the intra-observer variability. Table \ref{tab:reviewing-phase-variabilities} presents the mean intra- and inter-observer variabilities of both reviewers at the end of the reviewing phase. 

        \begin{table}[!th]
            \centering
            \caption{Mean inter- and intra-observer variabilities of senior orthodontists in terms of Mean Radial Error (MRE) $\pm$ Standard Deviation (SD) at the end of reviewing phase.}
            \begin{tabular}{cccc}
                \toprule
                {} & \multicolumn{2}{c}{Intra-observer Variability} & Inter-observer Variability \\ 
                \cmidrule(r){2-4}
                {} & Reviewer 1 & Reviewer 2 & Reviewer 1 vs. Reviewer 2 \\ 
                \midrule
                MRE (mm)  & {1.214 $\pm$ 1.150} & {1.348 $\pm$ 1.268} & {0.425 $\pm$ 0.552} \\ 
                \bottomrule
            \end{tabular}
            \label{tab:reviewing-phase-variabilities}
        \end{table}
        
\begin{figure}
            \centering
            \begin{tabular}{ccc}
                \subfloat[]{\includegraphics[width=5cm]{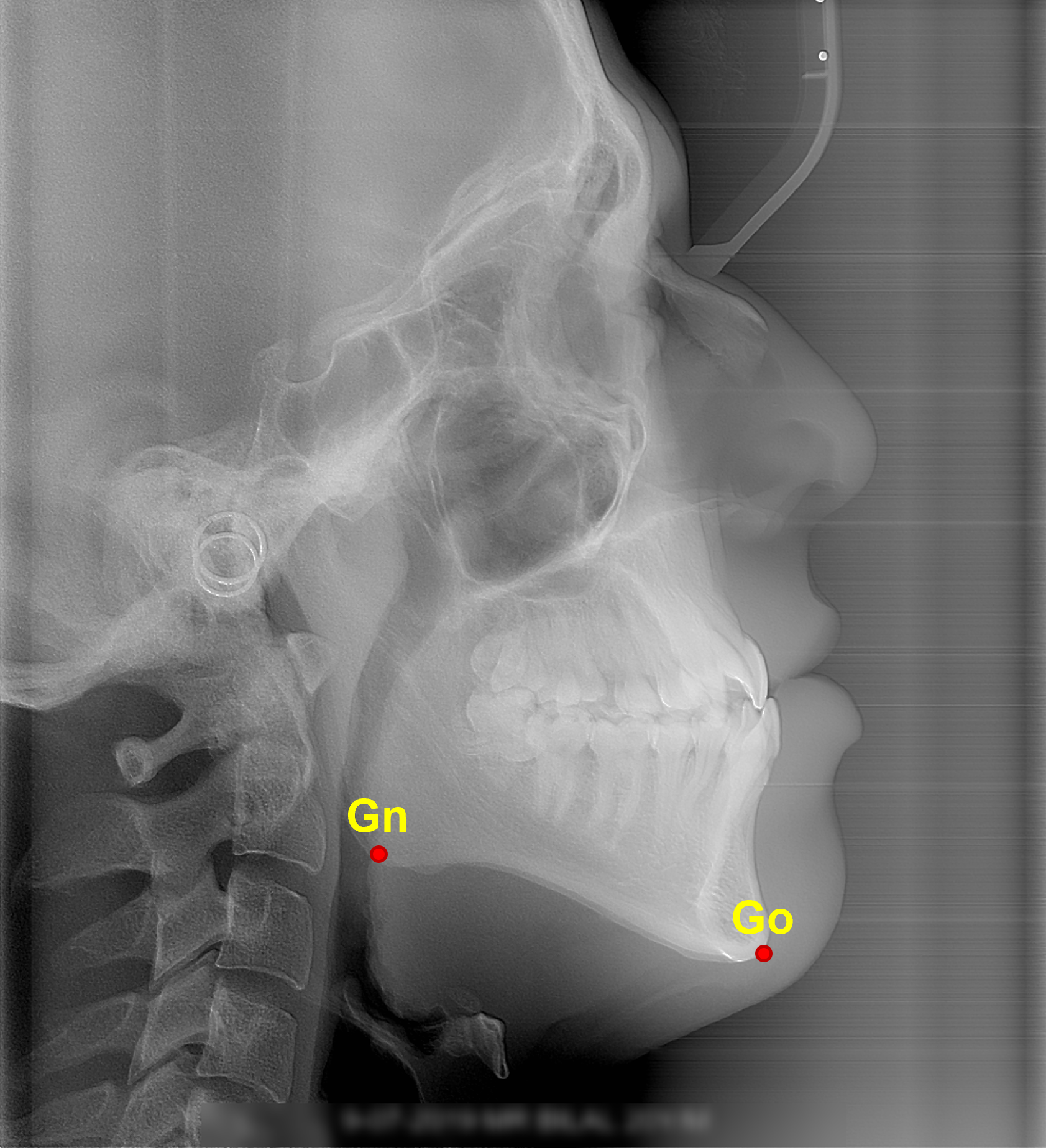}} &
                \subfloat[]{\includegraphics[width=5cm]{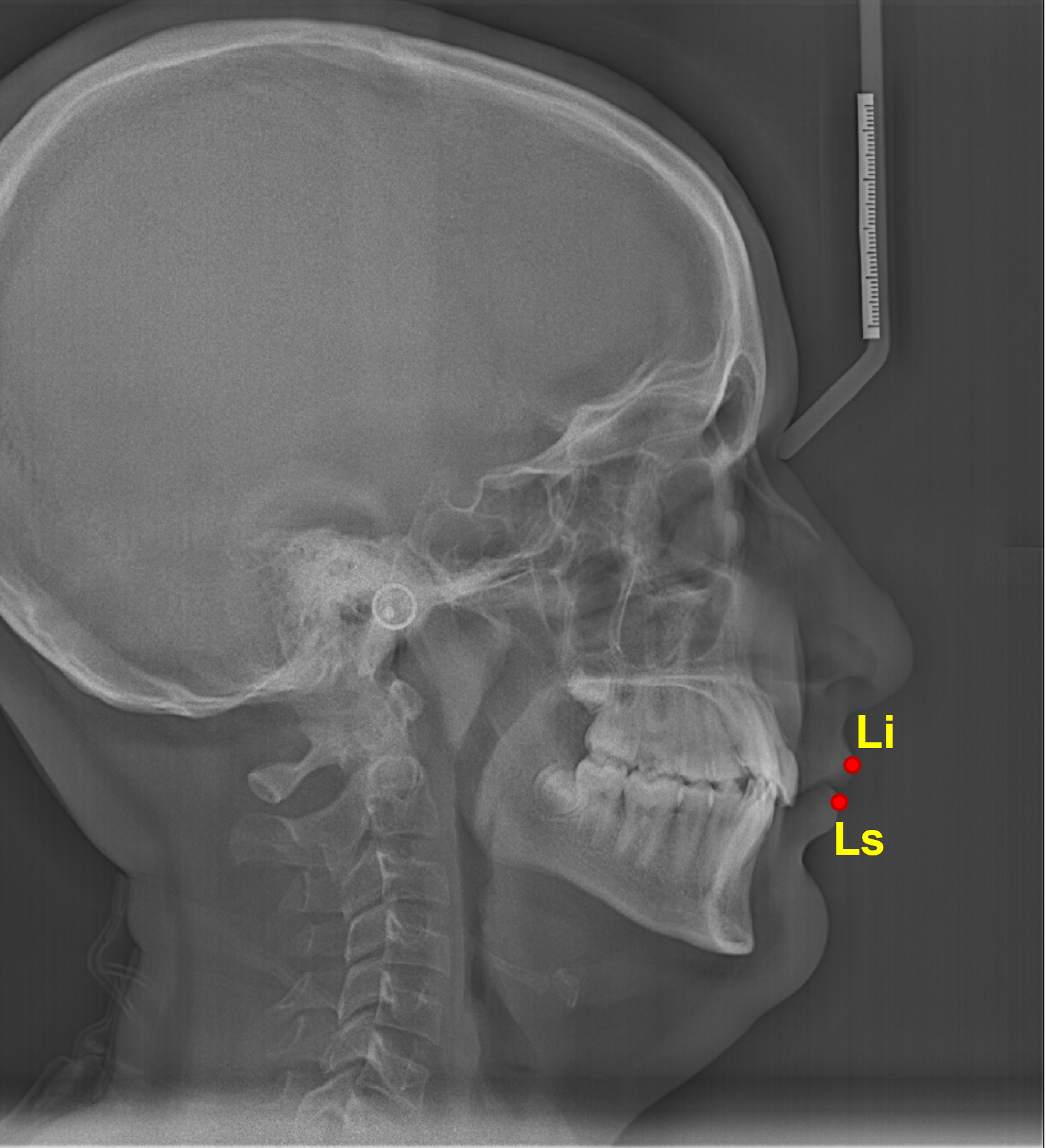}} & 
                \subfloat[]{\includegraphics[width=5cm]{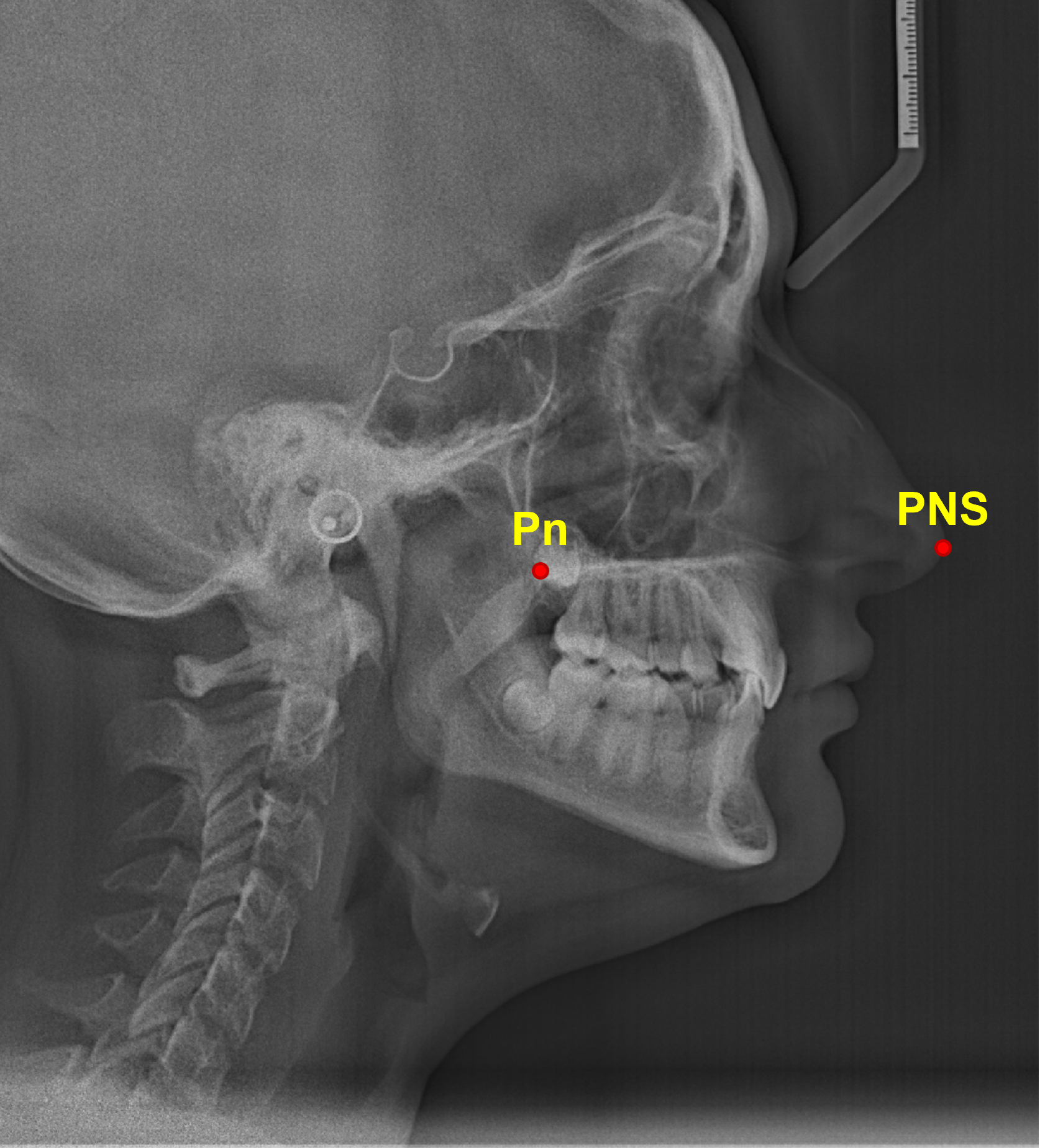}}
            \end{tabular}
            \caption{A visualization of the perils of mislabelling caused by swapped landmarks, exemplifying how two labellers accidentally swapped the locations of two landmarks, even though they were correctly marked in their respective positions. (a) shows the case where gonion (Go) was swapped with gnathion (Gn) (b) illustrates labrale inferius (Li) being switched with labrale superius (Ls) (c) presents posterior nasal spine (PNS) being swapped with pronasale (Pn)}
            \label{fig:swapped-landmarks-review-phase}
        \end{figure}

        Furthermore, during the reviewing phase, we identified a number of discrepancies in the annotations that could not be detected through statistical analysis alone. For instance, we observed a significant radial error between the markings when one of the labellers swapped two landmarks due to confusion between similar structures when annotating. Since the other labeller had correctly marked these landmarks, the statistical analysis flagged this error and it was corrected. However, such errors could have gone undetected if both labellers had made the same mistake. This was an alarming situation because the presence of such errors is a regular occurrence owing to human involvement, and we had already faced such a situation. As a result, we decided that our expert orthodontists will have to review the annotations of the entire dataset to ensure that the labellers have not marked any two landmarks in place of each other. Our diligence paid off, as this thorough review uncovered 8 instances where both the labellers had accidentally swapped the locations of two landmarks. We have included illustrations of some of these cases in Figure \ref{fig:swapped-landmarks-review-phase}.

    \subsubsection{CVM Stages}
    During the labelling process, our clinicians apprised the fact that there is no standard dataset available for automatic CVM stage classification. With this in mind, we decided to not only annotate cephalometric landmarks but also include CVM stage labels for each cephalogram in our dataset. However, identifying the CVM stage is a challenging task; in certain circumstances, even senior orthodontists have to put in considerable effort. Therefore, to ensure that the labellers were equipped with the necessary skills and knowledge to accurately identify CVM stages, the expert orthodontists conducted a comprehensive training session, primarily focused on the importance of following the standard CVM degree method in determining CVM stages. Unfortunately, after the completion of the labelling phase, we found that the CVM agreement between the two labellers was only 36.3\%.

    \begin{figure}[!ht]
        \centering
        \includegraphics[width=10.5cm]{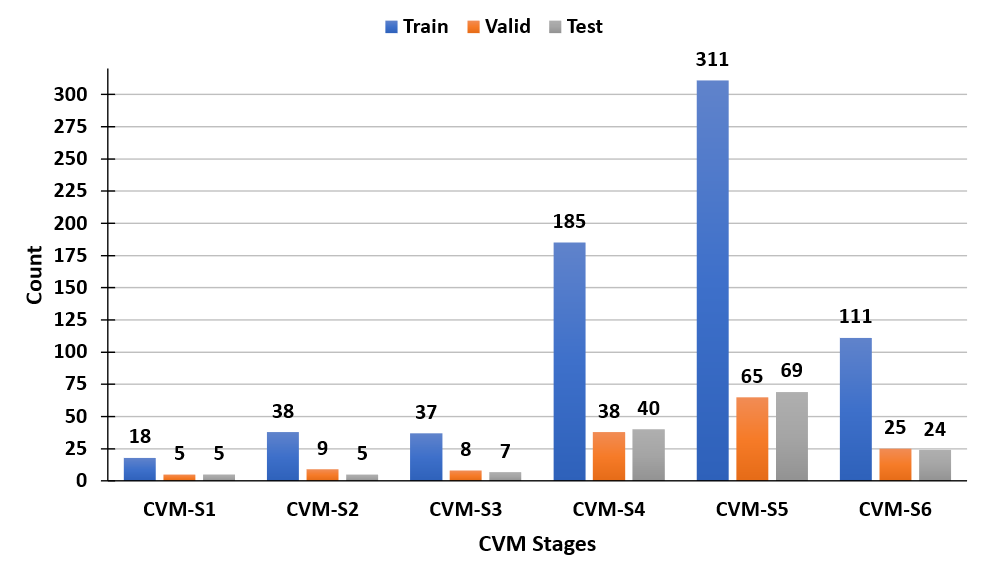}
        \caption{A visual breakdown of the distribution of CVM stages in our dataset}
        \label{fig:cvm-stages-bar-graph}
    \end{figure}

    Given the difficulties experienced by the junior orthodontists in identifying CVM stages, even with proper training, we decided to have the senior orthodontists take on the task of CVM stage labelling. Through their collaborative efforts, the agreement between the reviewers improved significantly to 96.6\%. of the 1000 cephalograms, the reviewers assigned the same label to 906 images, while differing in their assessment of the remaining 94. These 94 cephalograms were reviewed by expert orthodontists, and their markings were considered the final labels. A visual representation of the distribution of CVM stages in our dataset can be seen in Figure \ref{fig:cvm-stages-bar-graph}.
    
    \section{Usage Notes}    
    The dataset contains sensitive and valuable information that is protected by federal and state laws, which prohibit its unauthorized use or disclosure. Any individual with access to the dataset has a responsibility to comply with the laws and policies that govern such information. To ensure the privacy of patients in our dataset, we have implemented strict access controls and usage mechanisms. Researchers who wish to access the data must contact the corresponding author and provide a brief statement explaining their reasons for requesting access and their intended use of the data. We will review all applications on a case-by-case basis and grant access only to those who meet our criteria. Approved researchers will be required to sign and submit a Data Use and Confidentiality Agreement (DUA) before receiving access to the dataset. We ask all users of the dataset to adhere to our limitations on data use, including not using the data for commercial purposes or re-identification of individuals. Additionally, we encourage users to acknowledge the dataset in any publications or presentations resulting from the use of our data.
    
    \section{Code Availability}
    To facilitate the use of our dataset and to provide a starting point for researchers, we have made the source code for the data processing, transformations, and evaluation available on our GitHub repository\footnote{https://github.com/manwaarkhd/aariz-cephalometric-dataset}. The repository includes scripts for reading the cephalograms and corresponding annotations from their respective folders, as well as several transformations and augmentations that can be applied to the X-ray images, including contrast-limited adaptive histogram equalization \cite{reza2004realization}, unsharp masking \cite{malin1977unsharp} and histogram equalization \cite{pizer1987adaptive}. We have also provided scripts for the evaluation metrics to assess the performance of any AI algorithms applied to the dataset. We encourage researchers to use and modify our code as needed to adapt to their research questions and needs, and to provide feedback and suggestions for improvements.
    
    \section{Conclusion}
    The development of automated landmark detection systems has been hindered by a lack of reliable datasets. In an effort to address this gap, we present a novel dataset of lateral cephalometric radiographs (LCRs) annotated with 29 of the most commonly used anatomical landmarks, including 15 skeletal, 8 dental, and 6 soft-tissue landmarks. Our dataset, which comprises 1000 radiographs obtained from 7 different imaging devices with varying resolutions, is the most diverse and comprehensive cephalometric dataset to date. In addition to the extensive annotation of landmarks, our team of clinical experts also labelled the cervical vertebral maturation (CVM) stage of each radiograph, creating the first standard resource for CVM classification. Offering a diverse range of images acquired from different X-ray machines and providing a comprehensive set of annotations, our dataset has the potential to greatly improve the accuracy and reliability of automated cephalometric landmark detection systems, ultimately leading to more informed orthodontic treatment decisions.

    \section*{Acknowledgments}
    We would like to express our heartfelt gratitude to all of the clinicians at Islamic International Dental College for their tireless efforts and contributions to the annotation process of this dataset. This research would not have been possible without the support of Riphah International University, Islamabad, Pakistan. We also extend our sincere thanks to the patients who provided consent for the use of their cephalometric images. Finally, we would like to acknowledge the valuable feedback and suggestions provided by the anonymous reviewers, which helped us improve the quality of this dataset.
    

    \bibliographystyle{unsrt}  
    \bibliography{references}

    \begin{sidewaysfigure}[p]
        \includegraphics[width=23cm]{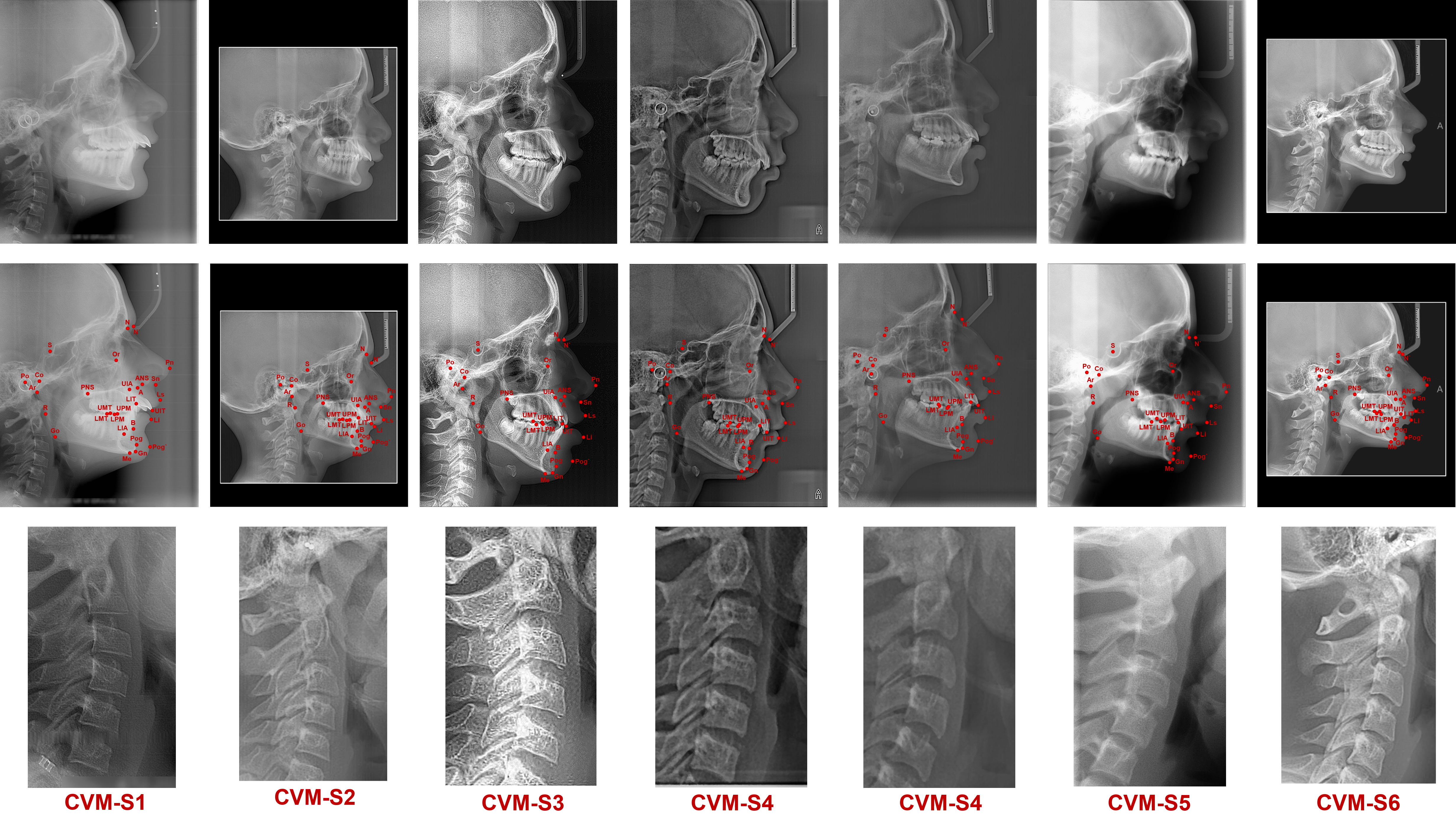}
        \caption{Seven shades of cephalograms: A diverse collection of sample images from various imaging devices along with their cephalometric landmarks and CVM stage}
        \label{fig:overall-dataset-records-visualization}
    \end{sidewaysfigure}
    
\end{document}